\documentclass[aps,prb,reprint,longbibliography]{revtex4-2}
\usepackage{amsmath}
\usepackage{amsfonts}
\usepackage{graphicx}
\usepackage{units}  
\usepackage{xspace}
\usepackage{subfigure}
\usepackage{hyperref}

\newcommand{\ve}{\varepsilon}
\newcommand{\mb}{\mathbf}
\newcommand{\mk}{\mathbf{k}}

\newcommand{\beq}{\begin{equation}}
\newcommand{\eeq}{\end{equation}}
\newcommand{\bea}{\begin{eqnarray}}
\newcommand{\eea}{\end{eqnarray}}
\newcommand{\mr}{\mathrm}

\usepackage[usenames]{color}

\begin{document}

\title{Effect of many-body interaction on de Haas-van Alphen oscillations in insulators}

\author{Gurpreet Singh and Hridis K. Pal}
	\affiliation{Department of Physics, Indian Institute of Technology Bombay, Powai, Mumbai 400076, India}
\date{\today}

\begin{abstract}
De Haas-van Alphen (dHvA) oscillations are oscillations in the magnetization as a function of the inverse magnetic field. These oscillations are usually considered to be a property of the Fermi surface and, hence, a metallic property. Recently, however, such oscillations have been shown to arise, both experimentally and theoretically, in certain insulators which have a narrow gap and an inverted band structure. In this work, we develop a theory to study the effect of many-body interaction on these unconventional oscillations. We consider weak interaction, focusing on the effect of renormalization of the quasiparticle spectrum on these oscillations. We find that interaction has an unusual effect: unlike in metals, in a certain regime the amplitude of oscillations may be enhanced substantially, both at zero and nonzero temperatures, even when the interaction is perturbatively weak.
\end{abstract}

\maketitle


One of the striking consequences of Landau quantization in a magnetic field in metals is the appearance of quantum oscillations. These are oscillations in physical observables of metals, both thermodynamic and transport, with a change in the magnetic field. The underlying mechanism is simple: as the magnetic field increases, the spacing between the Landau levels widens. This forces the highest occupied level to spill out of the Fermi level and become depopulated periodically, which manifests as oscillations in various observables. Evidently, these oscillations are expected only in metals and are measured routinely in experiments to map the Fermi surface of such systems \cite{de1930dependence,shoenberg_1984}. 

In recent years, however, quantum oscillations have been reported experimentally in various insulators. De Haas-van Alphen (dHvA) oscillations, which refer to oscillations in the magnetization, have been observed in the Kondo insulator SmB$_6$ \cite{li2014two,tan2015unconventional,hartstein2018fermi,liu2018fermi}, while  Shubnikov-de Haas (SdH) oscillations, which refer to oscillations in the resistivity, have been observed in the Kondo insulator YbB${_{12}}$ \cite{xiang2018quantum}, quantum well heterostructures \cite{PhysRevLett.123.126803,PhysRevLett.122.186802}, WTe$_2$ \cite{wang2021landau}, and moir{\'e} graphene \cite{liu2023quantum}. Such unexpected findings prompted intensive theoretical investigations which have now revealed that contrary to the canonical picture, quantum oscillations can indeed arise in insulators, provided the insulators have a narrow gap and an inverted band structure \cite{originalknolle,quanosczhang,chempotpal,fermiseapal,unusualpal}. 

Although the phenomenon is now well-understood at the noninteracting level, the effect of interactions on these unconventional oscillations remains insufficiently explored. Research in this direction has predominantly concentrated on specific models of correlated insulators where interaction causes the opening of the gap \cite{PhysRevLett.116.046403,PhysRevB.97.045152,PhysRevB.96.075115,PhysRevB.100.085124} but not on generic band insulators with interaction. Exploring the latter is important since it offers a controlled approach to compare oscillations in insulators with those in metals. Moreover, from an experimental perspective, investigating this aspect is significant since some of the systems where unconventional oscillations have been observed are of this nature \cite{PhysRevLett.123.126803,PhysRevLett.122.186802}, and the abundance of possibilities in this category suggests more explorations in the future.

In this work, we develop a theory of quantum oscillations in interacting band insulators, focusing specifically on dHvA oscillations. We consider weak interactions, incorporating it within the Hartree-Fock approximation.  Notably, we find that in a certain parameter regime, even weak interactions can significantly modify the amplitude of oscillations unlike in metals. This arises because interactions now competes with a new energy scale in the form of a gap which is absent in a metal, thus influencing oscillations in a qualitatively different manner.

To put our results for the insulator in context, we first review the theory of dHvA oscillations in metals. The oscillating part of the magnetization is given by $M_{\mr{osc}}=-\frac{\partial\Omega_{\mr{osc}}}{\partial B}$, where $\Omega_{\mr{osc}}$ is the oscillating part of the grand potential and $B$ is the external magnetic field. In a three-dimensional interacting metal with a parabolic spectrum of spinless electrons, considering only the renormalization of the energy spectrum due to a static interaction potential at the Hartree-Fock level, it is found that \cite{shoenberg_1984,lifshitz1956theory} ($\hbar=1$)
\beq
\tilde\Omega_{\mr{osc}}=\sum_{l=1}^{\infty}\tilde A_l(T)\mr{cos}\left[2\pi l\frac{\tilde\mu}{\tilde\omega_c}-\pi\left(l+\frac{1}{4}\right)\right],
\label{omegaoscfree}
\eeq
where $\tilde\omega_c=eB/\tilde m$ is the cyclotron frequency with $e$ as the absolute value of the charge and $\tilde m$ as the mass of an electron, respectively, $\tilde\mu$ is the chemical potential, and $\tilde A_l(T)$ is the temperature ($T$)-dependent amplitude of the $l$-th harmonic of oscillations given by ($k_B=1$)
\beq
\tilde A_l(T)=\tilde{\omega}_c\frac{(eB)^{3/2}}{8\pi^4 l^{5/2}}\frac{2\pi^2l T/\tilde\omega_c}{\mr{sinh}(2\pi^2 l T/\tilde\omega_c)}. 
\label{ampfree}
\eeq
Above, and henceforth, the presence (absence) of tilde denotes renormalized (bare) values of the respective quantities. The expressions~(\ref{omegaoscfree}) and (\ref{ampfree}) are, in fact, identical to the ones that appear in the noninteracting case, except for the appearance of renormalized parameters \cite{PhysRev.121.1251,wasserman}:
\begin{subequations}
\begin{align}
\mu&\rightarrow \tilde\mu=\mu(1+b) \\
\omega_c&\rightarrow\tilde\omega_c=\frac{eB}{\tilde{m}}=\frac{eB}{m}(1+a),
\end{align}
\label{metalrenorm}%
\end{subequations}
where $b$ and $a$ capture the degree of renormalization in $\mu$ and $m$, respectively. One can readily summarize the following salient features: 
\begin{enumerate}
\item The phase does not change with interaction.
\item The change in frequency is negligible: Because $\mu\gg\omega_c,V$, where $V$ is the strength of interaction, $\tilde{\mu}/\tilde\omega_c\approx\mu/\omega_c$.
\item At $T=0$, the change in the amplitude is small in proportion to the strength of the weak interaction since $\tilde A_l(0)\sim \tilde\omega_c$. The amplitude decreases monotonically with rise in $T$. 
\end{enumerate}

\begin{figure}
	\centering
	\includegraphics[width=0.6\columnwidth,height=45mm]{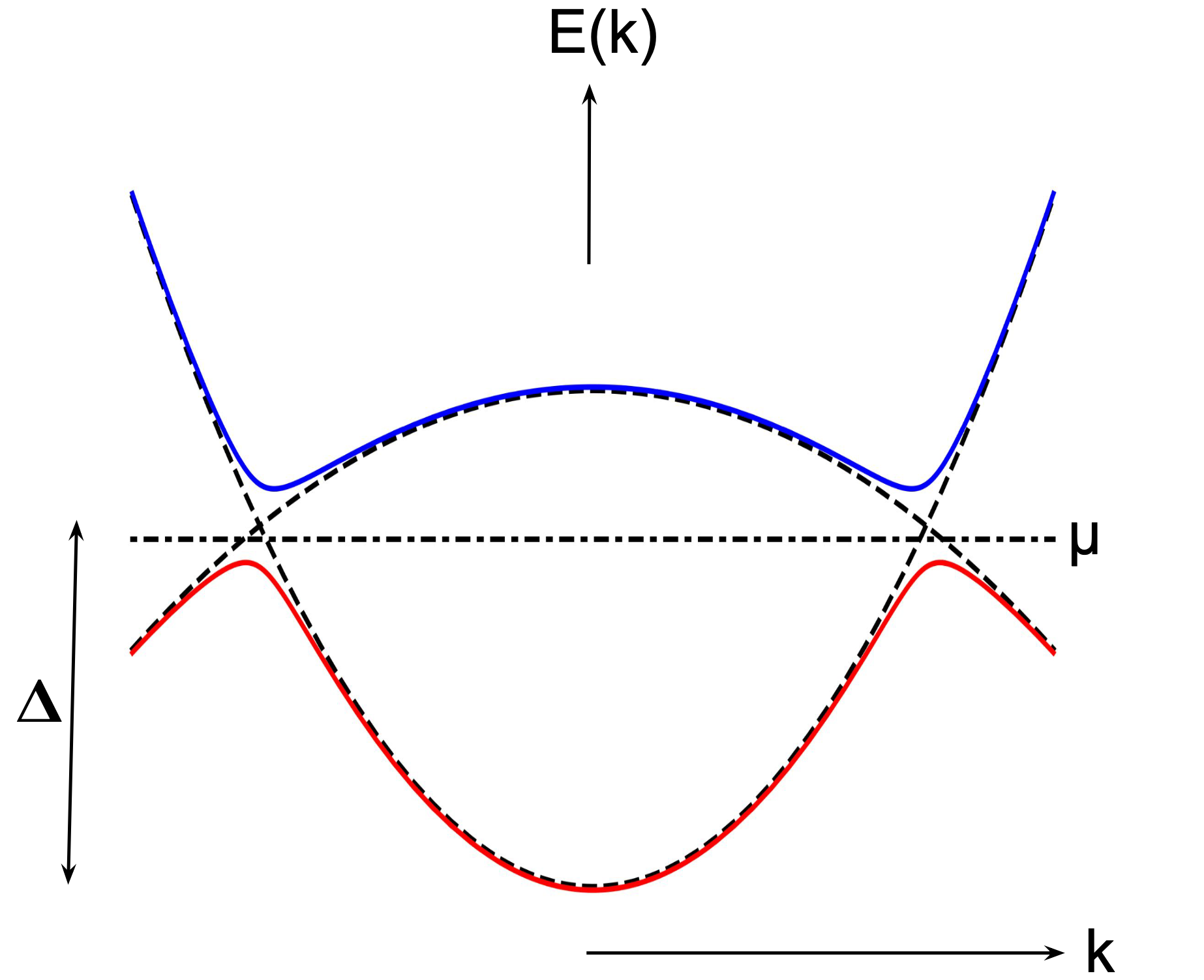}
	\caption{Schematic band structure for the Hamiltonian in (\ref{hamiltonian}) with $\epsilon_{1k}$ and $\epsilon_{2k}$ having curvatures of different sign. Dotted curves show bands before hybridization. The hybridization results in a gap.}
	\label{banddiag}
\end{figure}



We now proceed to investigate dHvA oscillations in an insulator. Consider the following Hamiltonian:
\begin{equation}
H=\sum_{i,\mk}\ve_{i\mk}c_{i\mk}^\dagger c_{i\mk}+\sum_{\mk}\left(\gamma_\mk c_{1\mk}^\dagger c_{2\mk}+\mr{h.c.}\right)+H_{\mr{int}}.
\label{hamiltonian}
\end{equation}
Here, $c_{i\mk}^\dagger (c_{i\mk})$, $i=1,2$, denotes the creation (destruction) operators for particles with momentum $\mk$ in the $i$-th band with dispersion $\ve_{i\mk}$ in three dimensions. These two bands are hybridized by $\gamma_\mk$. For simplicity, we choose $\ve_{1\mk}=\frac{k^2}{2m_1}-\Delta$ and $\ve_{2\mk}=-\frac{k^2}{2m_2}$ with $m_{1,2},\Delta >0$, and assume $\gamma_\mk=\gamma$ to be independent of $\mk$  with $|\gamma|\ll\Delta$. Also, all particles are assumed to be spinless. 
The first two terms in Eq.~(\ref{hamiltonian}) describe the noninteracting part and is easily diagonalized leading to an insulator with an inverted band structure and a narrow gap---see Fig.~\ref{banddiag}. The chemical potential $\mu$ is chosen to lie inside the gap. The last term in Eq.~(\ref{hamiltonian}), $H_{\mr{int}}$, introduces interparticle interaction, whose exact form is not necessary for the results to be derived---we only assume that the interaction is static and weak so that its effect can be included perturbatively at the Hartree-Fock level which leads to a renormalization of the energy levels but no broadening. 

\begin{figure*}
\centering
\includegraphics[width=\textwidth]{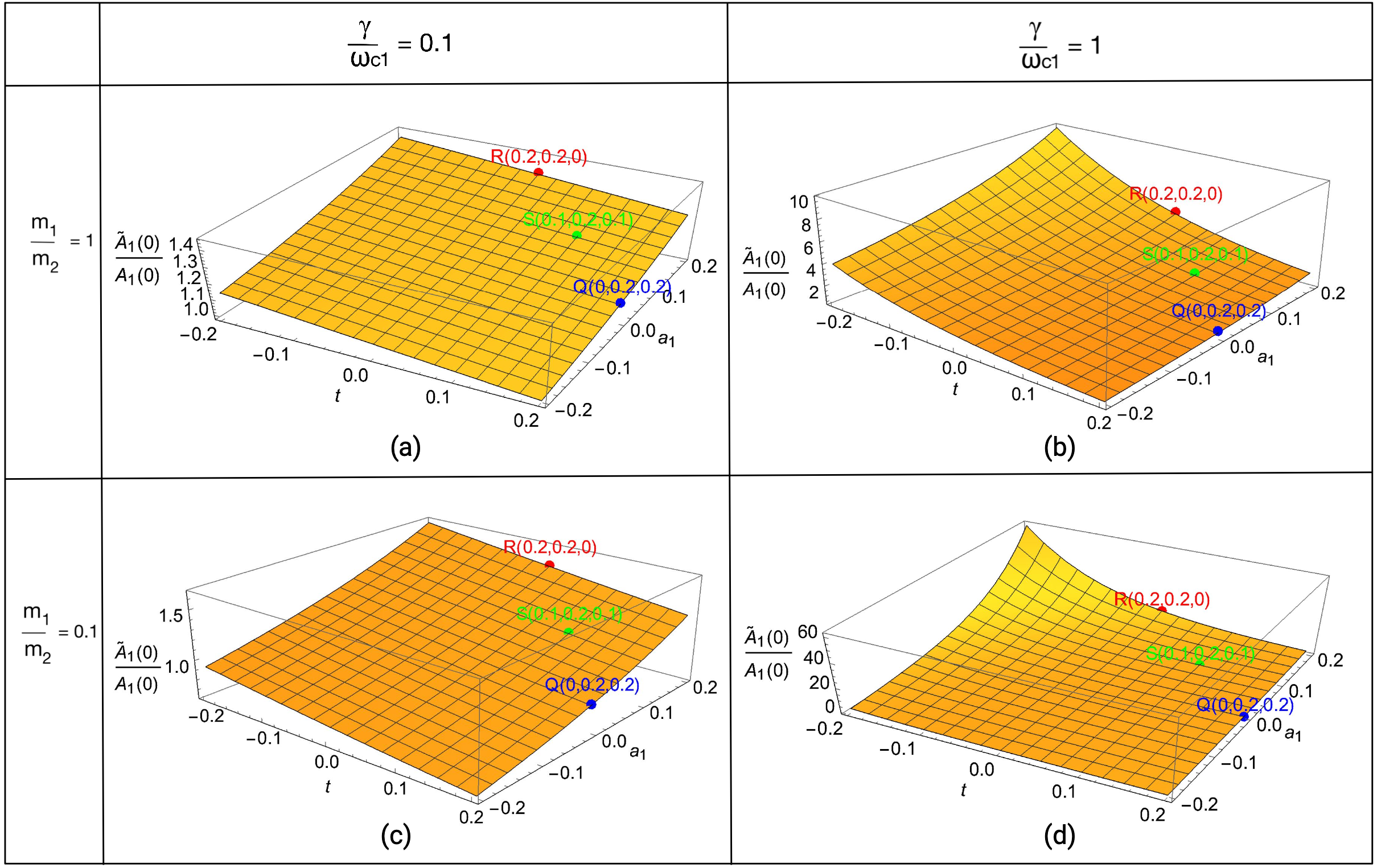}
\caption{The effect of interaction on the amplitude at $T=0$ according to Eqs.~(\ref{ampzero}) and (\ref{renorm}). $\tilde{A}_1(0)$ denotes the amplitude of the first harmonic at $T=0$ in the presence of interaction. It is normalized by its noninteracting value $A_1(0)$. We show its variation with $a_1$ and $t$ keeping $a_2=0.2$ fixed in all the plots. The points $Q, R$, and $S$ are arbitrarily chosen in the interaction-parameter-space defined by $(a_1,a_2,t)$ which are referred to in Fig.~\ref{amptemp} later. 
\label{ampint}}\end{figure*}


In the presence of a magnetic field, discrete Landau levels are produced that are affected by the interaction. We calculate the grand potential using the standard formula \cite{PhysRev.121.1251}:
\beq
\tilde\Omega=-T\ \text{Tr}\left(\sum_{{\zeta_m}}\ln\{-[\tilde{G}^{-1}(\zeta_m)]\}\right). 
\label{grand}
\eeq
Here,  $\xi_m=(2m+1)\pi T$, with $m\in\mathbb{Z}$, are the Matsubara frequencies, $\text{Tr}$ stands for the trace over all energy states, and $\tilde{G}$ is the Green's function corresponding to (\ref{hamiltonian}) in a magnetic field given by $\tilde{G}^{-1}=G^{-1}-\Sigma$, where $G$ is the noninteracting Green's function and $\Sigma$ is the self-energy due to the interaction. In the band-basis, we have  $G_{11}^{-1}=i\zeta_m-\omega_{c1}\left(n+\frac{1}{2}\right)-\frac{k^2_z}{2m_1}+\Delta+\mu$, $G_{22}^{-1}=i\zeta_m+\omega_{c2}\left(n+\frac{1}{2}\right)+\frac{k^2_z}{2m_2}+\mu$, and $G_{12}^{-1}=G_{21}^{-1}=-\gamma$, where $n$ is the Landau level index and $\omega_{c1,2}=eB/m_{1,2}$. In general, $\Sigma$ is a function of both $B$ and $T$ and requires a substantial effort to calculate. However, as far as dHvA oscillations in three dimensions are concerned, it suffices to consider $\Sigma$ calculated at zero $B$ and $T$---as in the case of metals, the effect of nonzero $B$ and $T$ leads to subleading corrections in orders of $\omega_{c1,2}/\Delta\ll 1$ and $T/\Delta\ll 1$, respectively \cite{wasserman,PhysRevB.73.045426,larsfritzgen}. Within this approximation, we evaluate the oscillating part of Eq.~(\ref{grand}) and find,
\beq
\tilde\Omega_{\mr{osc}}=\sum_{l=1}^{\infty}\tilde{A}_l(T)\mr{cos}\left[2\pi l\frac{\tilde\Delta}{\tilde\omega_{c1}+\tilde\omega_{c2}}-\pi\left(l+\frac{1}{4}\right)\right],
\label{omegagen}
\eeq
where
\begin{eqnarray}
\tilde{A}_l(T)&=&
\frac{(eB)^{3/2}}{\pi^2l^{3/2}}T\sum\limits_{\zeta_m>0}  e^{-\frac{\pi l  }{\tilde\omega_{c1}\tilde\omega_{c2}}\sqrt{\zeta_m^2(\tilde\omega_{c1}+\tilde\omega_{c2})^2+4\tilde\omega_{c1}\tilde\omega_{c2} \tilde\gamma^2}}\nonumber\\
&\times&\cosh\left[{\frac{\pi l\zeta_m (\tilde\omega_{c2}-\tilde\omega_{c1})}{\tilde\omega_{c1}\tilde\omega_{c2}}}\right].
\label{general} 
\end{eqnarray}
Details of the derivation are provided in Supplemental Material (SM) \cite{supp}. The above expressions are characterized by the following renormalized parameters:
\begin{subequations}
\begin{align}
\omega_{c1,2}&\rightarrow\Tilde{\omega}_{c1,2}=\frac{eB}{\tilde{m}_{1,2}}=\frac{eB}{m_{1,2}}(1+a_{1,2}),\\
\Delta&\rightarrow\tilde\Delta=\Delta (1+b),\\
\gamma&\rightarrow\tilde\gamma=\gamma(1+t),\\
\mu&\rightarrow\tilde\mu=\mu+\delta\mu.
\end{align}
\label{renorm}%
\end{subequations}
Note that, while $\tilde\omega_{c1,2}$, $\tilde\Delta$, and $\tilde\gamma$ appear explicitly in Eqs.~(\ref{omegagen}) and (\ref{general}), $\tilde\mu$ enters implicitly through the Matsubara sum. Thus, it affects only the T-dependence of the oscillations, and has no effect on the $T=0$ behavior, as long as $\mu$ and $\tilde\mu$ lie in the gap. For simplicity, we have assumed $\tilde\mu=-\frac{\tilde\Delta \tilde m_1}{\tilde m_1+\tilde m_2}$, chosen such that it lies at the intersection of the two renormalized bands prior to hybridization \footnote{We assume that $\mu$ is a free parameter chosen such that it results in the $\tilde\mu$ assumed. This is not a unique choice---the chemical potential can lie anywhere inside the gap. Although this has no consequence at $T=0$, new qualitative features can arise at $T\ne 0$. This was demonstrated in Ref.~\cite{chempotpal} for the noninteracting case; additional new features may be expected in the interacting case which we do not consider in this work.}. Equation~(\ref{omegagen}) along with Eqs.~(\ref{general}) and (\ref{renorm}) generalize Eqs.~(\ref{omegaoscfree}), (\ref{ampfree}), and (\ref{metalrenorm}) from an interacting metallic system to an interacting gapped system. Thus, for a given form of $H_{\mr{int}}$ in Eq.~(\ref{hamiltonian}), one simply needs to calculate the renormalization parameters $a_i,b,t$ to study the effect of interaction on dHvA oscillations. We will come back to this calculation later; for now, we discuss the qualitative features that arise from these expressions. It is seen that the phase remains unaltered and the frequency does not change appreciably since $1\ll\frac{\tilde\Delta}{\tilde\omega_{c1}+\tilde\omega_{c2}}\approx\frac{\Delta}{\omega_{c1}+\omega_{c2}}$; thus, both these quantities remain qualitatively similar to those in metals. In contrast, the amplitude is significantly affected by interactions, in a way that is qualitatively different from that in metals.

First, we consider $\tilde A_l(0)$, the amplitude at $T=0$. Changing the summation to an integral over the frequency in Eq.~(\ref{general}), we find, 
\beq
\tilde A_l(0)=\frac{|\tilde\gamma|(eB)^{3/2}}{2\pi^3 l^{3/2}}K_1\left(\frac{4\pi l|\tilde\gamma|}{\tilde\omega_{c1}+\tilde\omega_{c2}}\right),
\label{ampzero}
\eeq
where $K_\alpha$ is the modified Bessel function of the second kind. Equation (\ref{ampzero}), together with Eqs.~(\ref{renorm}), gives a quantitative description of how the amplitude is affected by interaction. It leads to an unusual feature that is unique to the insulating case: even a weak interaction can lead to a substantial change in the amplitude of oscillations at zero temperature. Indeed, the fate is decided by a delicate interplay between $\tilde{\gamma}$ and $\tilde{m}_{1,2}$ in Eq.~(\ref{ampzero}). Using the parametrization of Eq.~(\ref{renorm}) in Eq.~(\ref{ampzero}), we plot the first-harmonic-amplitude in Fig.~\ref{ampint} for different values of $m_1/m_2$ (determining the particle-hole asymmetry) and $\gamma/\omega_{c1}$ (determining the strength of the gap as compared to the Landau level spacing). It is seen that when $\gamma/\omega_{c1}\sim 1$, there is a pronounced enhancement in the zero-temperature-amplitude, which can amount to even an order of magnitude.
\begin{figure*}
	\centering
		\includegraphics[width=\textwidth]{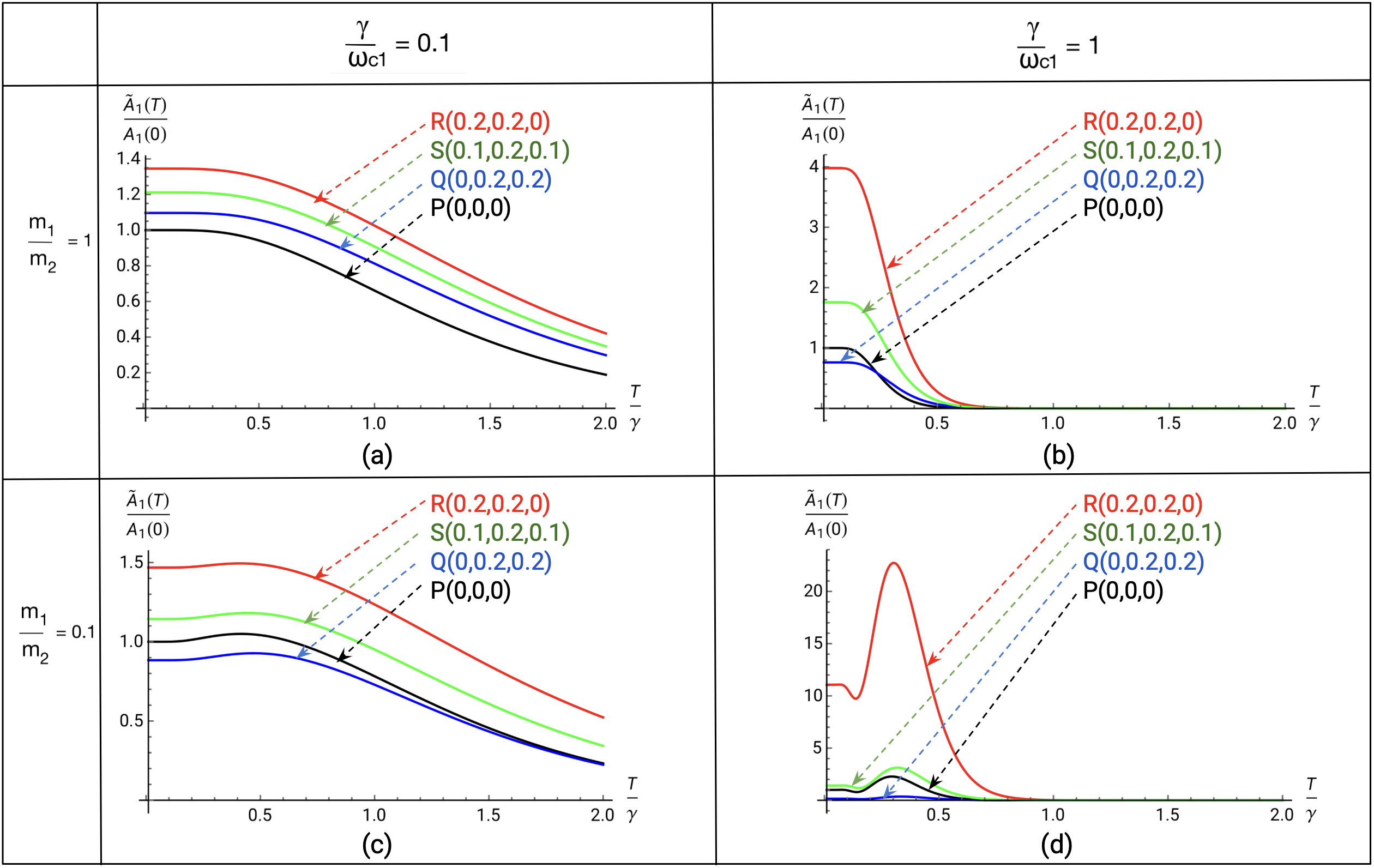}
	\caption{The temperature-dependence of the amplitude in the presence of interaction. $\tilde{A}_1(T)$ denotes the amplitude of the first harmonic at temperature $T$ in the presence of interaction. It is normalized by its noninteracting zero-temperature value $A_1(0)$. We show its variation with $T$ for the various cases considered in Fig.~\ref{ampint}.  $Q, R, S$ denote three choices of $(a_1,a_2,t)$ defined in Fig.~\ref{ampint} earlier. $P$ denotes the noninteracting case. The curves are derived from numerical calculation of Eq.~(\ref{general}).}\label{amptemp}
\end{figure*}


Next, we consider the dependence of the amplitude on $T$. This is calculated numerically from Eq.~(\ref{general}) and is presented in Fig.~\ref{amptemp} for various choices of interaction parameters $m_1/m_2$ and $\gamma/\omega_{c1}$ as used in Fig.~\ref{ampint}. In the limit $T\gg\gamma$, as expected, the $T$-dependence follows the usual metallic behavior, contributed by the two participating bands. It only depends on $\tilde m_{1,2}$ and is independent of $\tilde\gamma$. In the other limit, $T\lesssim\gamma$, the $T$-dependence deviates from the metallic behavior, and depends on both $\tilde m_{1,2}$ and $\tilde\gamma$. The deviation is most striking when $m_1/m_2\gg 1$ and $\gamma/\omega_{c1}\sim 1$. In this regime [Fig.~\ref{amptemp}(d)] There is a sizeable enhancement in the amplitude in the form of a nonmonotonic upturn driven by $T$ on top of the enhancement at $T=0$ discussed earlier. An interesting observation is that since the behavior of the amplitude at low $T$ in the particle-hole asymmetric case is very sensitive to $\gamma/\omega_{c1}$, for a given material (with a fixed $\gamma$ and $\tilde{\gamma}$) the effect of temperature depends crucially on the field at which the oscillations are being studied to extract the amplitude. Indeed, Figs.~\ref{amptemp}(c) and (d) can be interpreted as the temperature dependence of the amplitude of the same dHvA oscillation but at different field values. 

An expression for the $T-$dependence of the amplitude at low $T$ can be obtained by employing the Euler-Maclaurin formula to carry out the Matsubara sum in Eq.~(\ref{general}):
\beq
\tilde{A}_k(T)\approx\tilde{A}_k(0)-\alpha\left[\frac{1}{2\pi} \int\displaylimits_{0}^{\pi T} f(x) dx -\frac{T}{2}f(\pi T)+\frac{T}{12}f'(\pi T)\right],\label{amptana}
\eeq
where $f(x)$ is the summand in Eq.~(\ref{general}) with $\zeta_m\rightarrow x$ and $\alpha=\frac{(eB)^{3/2}}{\pi^2l^{3/2}}$. Equation (\ref{amptana}) reproduces the  numerically obtained curve for $\tilde{A}_1(T)$ very well for $T\lesssim\gamma$. This can be further reduced to a closed analytical form by expanding in $T$; however, the resulting expression is accurate only when $T\ll\gamma$ and does not describe the features that arise at $T\lesssim\gamma$. We discuss this in SM \cite{supp}.


Having discussed the qualitative features, we now return to the calculation of the renormalization parameters $a_i,b,t$. These are related to the self-energy $\Sigma$ as follows (see SM \cite{supp}):
\begin{subequations}
\begin{align}
a_i&=v_{Fi}^{-1}\partial_k \Sigma_{ii}\big\vert_{k_F},\\
b&=\frac{1}{\Delta}\left[\Sigma_{22}-\Sigma_{11}+\mu\left(a_1-a_2\right)\right]\big\vert_{k_F},\\
t&=\frac{1}{\gamma}\Sigma_{12}\big\vert_{k_F},\\
\delta\mu&=\mu a_2-\Sigma_{22}\big\vert_{k_F},
\end{align}
\label{renormcalc}%
\end{subequations}
where $v_F$ and $k_F$ are the Fermi velocity and momentum, respectively. To calculate $\Sigma$, we need $H_{\mr{int}}$. We consider the following form:
\beq
H_{\mr{int}}=\sum_{i,j,\mk,\mk^\prime,\mb{q}}V_{ij\mb{q}}	c^\dagger_{i\mk+\mb{q}}c^\dagger_{j\mk^{\prime}-\mb{q}}c_{j\mk^{\prime}}c_{i\mk}.
\eeq
The above term describes an interaction of strength $V_{ij\mathbf{q}}$ between particles belonging to either the same band ($i=j$) or different bands ($i\ne j$). For simplicity, we have only considered interband interaction which preserves the band index, but extending the calculation to a more general form of the interaction is straightforward. As remarked earlier, $\Sigma$ needs to be calculated at zero $B$ and $T$. At the Hartree-Fock level, we have $\Sigma_{ii}(\mb{k})=\sum_{\mb{k}'}\left[\left(V_{ii\mb{0}}-V_{ii\mb{k}'-\mb{k}}\right)\langle c_{i\mb{k}'}^{\dagger}c_{i\mb{k}'}\rangle+V_{12\mb{0}}\langle c_{j\mb{k}'}^{\dagger}c_{j\mb{k}'}\rangle\right]$ ($j\ne i$) and $\Sigma_{12}(\mb{k})=-\sum_{\mb{k}'}V_{12\mb{k}'-\mb{k}}\langle c_{1\mb{k}'}^{\dagger}c_{2\mb{k}'}\rangle$. The averages over the ground state at $T=0$, denoted by $\langle\cdots\rangle$, are easily computed (see SM \cite{supp}). Finally, we obtain,
\begin{subequations}
\begin{align}
\Sigma_{ii}(\mb{k})&\approx\sum_{\mb{k}'}\left[\left(V_{ii\mb{0}}-V_{ii\mb{k}'-\mb{k}}\right)n_F(\xi_{i\mb{k}'})+V_{12\mb{0}}n_F(\xi_{j\mb{k}'})\right],\\
\Sigma_{12}(\mb{k})&=\gamma\sum_{\mb{k}'}V_{12\mb{k}'-\mb{k}}\frac{n_F(E_{-\mb{k}'})}{(E_{+\mb{k}'}-E_{-\mb{k}'})},
\end{align}
\label{selfenexp}%
\end{subequations} 
where $\xi_{i\mb{k}'}=\ve_{i\mb{k}'}-\mu$, $n_F(x)$ is the Fermi function, and $E_{\pm\mb{k}'}=\frac{1}{2}\Big[(\ve_{1\mb{k}'}+\ve_{2\mb{k}'})\pm\sqrt{(\ve_{1\mb{k}'}-\ve_{2\mb{k}'})^2+4\gamma^2}\Big]$.  In deriving $\Sigma_{ii}$, we have neglected terms of $\mathcal{O}(\gamma^2/\Delta^2)$.

\begin{figure}
	\centering
	\includegraphics[width=\columnwidth]{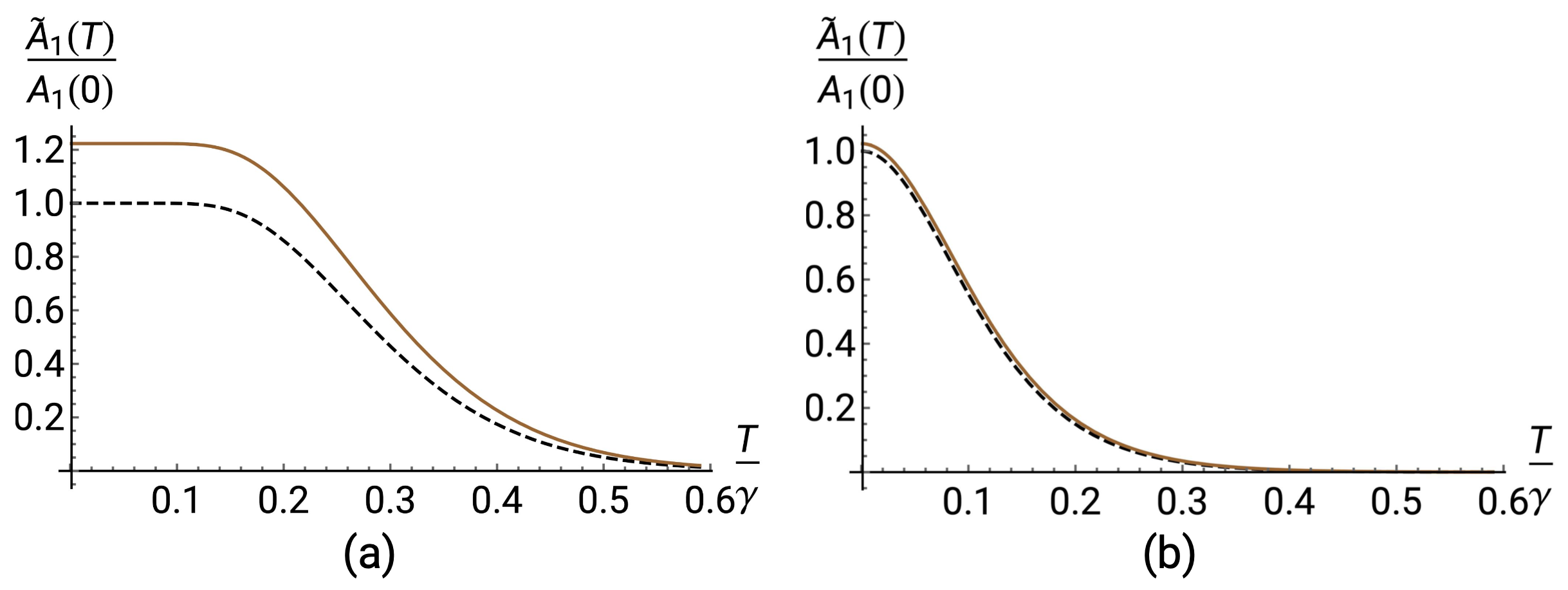}
	\caption{Comparison of amplitude between (a) an insulator and (b) a metal with (solid) and without (dashed) interactions. The amplitude is calculated using Eq.~(\ref{general}) and is normalized by its noninteracting value at $T=0$, We have used $m_1=m_2=m$ (electron's mass), $B=1$ T, and $\frac{\Delta}{\omega_c}= 4.3\times10^5 $. In (a) $\frac{\gamma}{\omega_c}= 0.9$ and in (b) $\gamma=0$ [$T$ is scaled with the $\gamma$ used in (a) for comparison]. The interaction term is taken as $V_{ij\mb{q}}=\frac{e^2}{\epsilon_0 (q^2+\kappa^2)}$, where $\kappa$ is the Thomas-Fermi wavevector. This yields $a_1=a_2=0.022$ and $t=0.329$ in Eqs.~(\ref{renormcalc}).}
	\label{metalinsulator}
\end{figure}
 
The amplitude of oscillations is determined by $\tilde\omega_{c1,2}$ and $\tilde\gamma$, which depend on $a_i$ and $t$, respectively. Using Eqs.~(\ref{selfenexp}) in ~(\ref{renormcalc}), we make an important observation: the two parameters depend on interaction in qualitatively different ways. While $t$ is proportional to the interaction potential, $a_i$ is proportional to its derivative. Thus, while $t$ relies solely on the interaction strength, $a_i$ is influenced by the momentum dependence of the interaction, in addition to its strength.  As a consequence, generic interactions are more likely to influence oscillations in an insulator by renormalizing the gap rather than the mass, and oscillations in insulators are more susceptible to interactions compared to metals. For instance, if one assumes a contact potential with $V_{ij\mb{q}}=U$, a constant, $a_i$ vanishes but $t$ is nonzero. Consequently, oscillations in a metal would remain unchanged, whereas they would be affected in an insulator, highlighting a key distinction between metals and insulators. As another example, we consider a particle-hole symmetric insulator interacting via the Thomas-Fermi screened Coulomb interaction  $V_{ij\mb{q}}=\frac{e^2}{\epsilon_0 (q^2+\kappa^2)}$, where $\kappa$ is the Thomas-Fermi wavevector and $\epsilon_0$ is the electrical permittivity of free space. We plot the amplitude as a function of $T$ for a particle-hole symmetric case ($m_1=m_2$) in Fig.~\ref{metalinsulator}, comparing the insulating case with its metallic counterpart. The change in amplitude due to interaction is considerably more pronounced in the insulating case. 


In summary, we observe the following salient features of dHvA oscillations in interacting insulators: 
\begin{enumerate}
\item The phase does not change with interaction. 
\item The change in frequency is negligible: Because $\Delta\gg\omega_{c1,2},V$, where $V$ is the strength of interaction, $\tilde{\Delta}/(\tilde\omega_{c1}+\tilde\omega_{c2})\approx\Delta/(\omega_{c1}+\omega_{c2})$.
\item The change in the amplitude can be substantial even in weak interaction, at both $T=0$ and $T\ne 0$, in a certain regime. The amplitude may vary nonmonotonically with $T$ showing an upturn at low $T$, which is greatly amplified by interactions.
\end{enumerate}
In comparing the above features with their counterpart for metals stated earlier, we find that, while the phase and the frequency behave similarly, the behavior of the amplitude is very different, both qualitatively and quantitatively. 

In this work, we have considered weak interaction at the Hartree-Fock level. Going beyond, one can include the effects of a finite lifetime induced by interactions or disorder, which is expected to give rise to other features distinct from that in metals. We defer these questions for future investigation. Note, however, the assumption of weak interaction does not necessarily imply that our results do not apply to correlated insulators, such as Kondo and excitonic insulators \cite{continentino1995excitonic,allocca}, where interaction is strong. In such systems, the strong interaction is primarily responsible for giving rise to the insulating gap via a phase transition. Once such a phase is reached, the effective interaction between the new quasiparticles may indeed be weak. In passing, we note that in the Kondo insulator $\mr{SmB}_6$, the dHvA oscillations show an unusual enhancement of the amplitude at low $T$ \cite{tan2015unconventional,hartstein2018fermi,liu2018fermi}. This material is strongly particle-hole asymmetric and is most likely in the regime of Fig.~\ref{amptemp}(d). The upturn observed in $\mr{SmB}_6$ experimentally (see, for example, Fig. 5c in Ref.~\onlinecite{liu2018fermi}) bears some resemblance to the upturn in Fig.~\ref{amptemp}(d).


In conclusion, we have presented a theory to study the effect of many-body interactions on dHvA oscillations in insulators at the Hartree-Fock level. We have shown that the amplitude of oscillations can change substantially even if the interaction is weak, unlike in metals. The difference originates from the interplay between interactions and the gap in the band structure, a feature absent in metals. 


\begin{acknowledgments}
GS would like to thank CSIR for financial support via JRF. HKP would like to thank IRCC, IIT Bombay for financial support via grant RD/0518-IRCCSH0-029.
\end{acknowledgments}

\begin{widetext}
\section*{Supplemental Material}

\subsection{THE OSCILLATORY GRAND POTENTIAL}

We want to calculate the oscillatory part of the Grand potential given by
\beq
\tilde\Omega=-T \ \text{Tr}\left(\sum_{{\zeta_m}}\ln\{-[\tilde{G}^{-1}(\zeta_m)]\}\right),
\label{grand}
\eeq
where
\begin{equation}\tilde{G}^{-1}(\zeta_m)=\begin{bmatrix}
		i\zeta_m-\omega_{c1}\left(n+\frac{1}{2}\right)-\frac{k^2_z}{2{m}_1}+\Delta+\mu&-\gamma \\ -\gamma &i\zeta_m+\omega_{c2}\left(n+\frac{1}{2}\right)+\frac{k^2_z}{2{m}_2}+\mu
	\end{bmatrix}-\Sigma,\label{gdef}\end{equation}
 Here,  $\xi_m=(2m+1)\pi T$, with $m\in\mathbb{Z}$, $n$ is the Landau level index, $\omega_{c1,2}=eB/m_{1,2}$, and $\text{Tr}$ stands for the trace over all energy states, i.e., $ \text{Tr}=\frac{eB}{4\pi^2}\sum_{n}\int_{-\infty}^{\infty}dk_z$. Using the well-known relation Tr(ln M)=Tr[ln$|M|$], where $|M|$ is the determinant of any matrix M, we get
\beq
\tilde\Omega=-D T\int_{-\infty}^{\infty}d{k_z}\sum_{n}\left(\sum_{{\zeta_m}}\ln\{-|\tilde{G}^{-1}(\zeta_m,n,k_z)|\}\right),
\label{grand1}
\eeq
where $D=\frac{eB}{4\pi^2}$ and we have explicitly written all the variables on which $\tilde{G}^{-1}$ depends for clarity. The summation over the Landau levels in Eq.~(\ref{grand1}) can be changed into an integral with the help of Poisson's summation formula:
$\sum_{n}F(n)=\int_{0}^{\infty} F(x) dx + 2 \sum_{l=1}^{\infty}\int_{0}^{\infty}dx F(x) \cos(2\pi lx)$. It is evident that only the second term induces oscillations. To evaluate this term, we use the method described in Refs.~\onlinecite{PhysRevB.73.045426,larsfritzgen}. Utilizing the method of integration by parts and retaining only the terms that contribute to oscillations, we obtain:
\begin{equation} \tilde\Omega_{osc}=2DT \int_{-\infty}^{\infty}d{k_z}\left(\sum_{l=1}^{\infty}\sum_{{\zeta_m}}\int_{0}^{\infty}   \frac{1}{|-\tilde{G}^{-1}(\zeta_m,x,k_z)|}\frac{d}{dx}{|-\tilde{G}^{-1}(\zeta_m,x,k_z)|\frac{\sin{2\pi lx}}{2\pi l}dx}\right).  \label{eq}\end{equation}
It can be seen that the discrete variable $n$ has been replaced by a continuous variable $x$ above. To proceed further, we need to know how $\Sigma$ depends on the three variables $\zeta_m,\ x,\ k_z$ in Eq.~(\ref{gdef}). First, we assume a static interaction so that $\Sigma$ has no dependence on $\zeta_m$, i.e., $\Sigma(\zeta_m,x,k_z)\rightarrow \Sigma(x,k_z)$.
Next, for any generic weak interaction, one can expand the self-energy near $x=x_F$ and $k_z=0$ \cite{PhysRevB.73.045426,larsfritzgen}. Writing explicitly in the band-basis,
\begin{subequations}
\begin{align}
    \Sigma_{ii}(x,k_z)&\approx\Sigma_{ii}(x_F,0)+(x-x_F)\frac{\partial\Sigma_{ii}(x,k_z)}{\partial x}\Bigg|_{x_F,0}+\frac{k_z^2}{2}\frac{\partial^2\Sigma_{ii}(x,k_z)}{\partial k_z^2}\Bigg|_{x_F,0},\\
   \Sigma_{12}(x,k_z)&\approx\Sigma_{12}(x_F,0).
\end{align}
\label{sigma}%
\end{subequations}
Above, $x_F=(\mu+\Delta)/\omega_{c1}-1/2=-\mu/\omega_{c2}-1/2$.  The expansions above have been carried out to the leading order which affects oscillations. For brevity, we write $\frac{1}{\omega_{ci}}\frac{\partial\Sigma_{ii}(x,k_z)}{\partial x}\Bigg|_{x_F,0}=a_i$ and $\frac{\partial^2\Sigma_{ii}(x,k_z)}{\partial k_z^2}\Bigg|_{x_F,0}=\frac{1}{m_i}\alpha_i$.  A considerable simplification occurs by noting that it suffices to calculate $\Sigma_{ij}(x_F,0)$, $a_i$, and $\alpha_i$ at zero field since the effect of the field is to introduce higher order corrections in the field in oscillations which are negligible \cite{PhysRevB.73.045426,larsfritzgen,wasserman}. Further, it is also easily shown that $a_i=\alpha_i$ at zero field. With this in mind, we define the following:
 \begin{subequations}
\begin{align}
\Tilde{\omega}_{c1,c2}&=\omega_{c1,c2}(1+a_{1,2}),\\
\tilde\Delta&=\Delta (1+b),\\
\tilde\gamma&=\gamma(1+t),\\
\tilde\mu&=\mu+\delta\mu,
\end{align}
\label{renorm}%
\end{subequations}
where
\begin{subequations}
\begin{align}
a_i&= v_{Fi}^{-1}\partial_k \Sigma_{ii}\big\vert_{k_F},\\
b&=\frac{1}{\Delta}\left[\Sigma_{22}-\Sigma_{11}\right]\big\vert_{k_F}+\frac{1}{\Delta}\mu\left(a_1-a_2\right),\\
t&=\frac{1}{\gamma}\Sigma_{12}\big\vert_{k_F},\\
\delta\mu&=\mu a_2-\Sigma_{22}\big\vert_{k_F},
\end{align}
\label{renormcalc}%
\end{subequations}
where $v_F$ and $k_F$ are the Fermi velocity and momentum, respectively. Using (\ref{sigma}) in (\ref{gdef}) along with the definitions (\ref{renormcalc}) and (\ref{renorm}), we have
\begin{equation}\tilde{G}(\zeta_m)^{-1}=\begin{bmatrix}
		i\zeta_m-\tilde{\omega}_{c1}\left(x+\frac{1}{2}\right)-\frac{k^2_z}{2\tilde{m}_1}+\tilde{\Delta}+\tilde{\mu}&-\tilde{\gamma}\\ -\tilde{\gamma}&i\zeta_m+\tilde{\omega}_{c2}\left(x+\frac{1}{2}\right)+\frac{k^2_z}{2\tilde{m}_2}+\tilde{\mu}
	\end{bmatrix}.\end{equation}
We now solve Eq.~(\ref{eq}) using the above expression for $\tilde G^{-1}$. The determinant of matrix $-\tilde{G}^{-1}$ is the product of its eigenvalues. Setting $\tilde{\mu}$ at $\tilde{\mu}=-\frac{\tilde{\omega}_{c2}\tilde{\Delta}}{\tilde{\omega}_{c1}+\tilde{\omega}_{c2}}$ and after solving, we get
\begin{equation}\label{omega4}\tilde{\Omega}_{osc}=\frac{DT}{\pi} \int_{-\infty}^{\infty}d{k_z}\left(\sum_{l=1}^{\infty}\frac{1}{l}\sum_{{\zeta_m}}\left(-\frac{1}{\tilde{\omega}_{c1}\tilde{\omega}_{c2}}\right)\int_{0}^{\infty}\frac{\sin{2\pi lx}\left(i\zeta_m(\tilde{\omega}_{c2}-\tilde{\omega}_{c1})+\frac{2\tilde{\omega}_{c1}\tilde{\omega}_{c2}}{\tilde{\omega}_{c1}+\tilde{\omega}_{c2}}\tilde\Delta-2\tilde{\omega}_{c1}\tilde{\omega}_{c2} x-(1+\frac{k^2_z}{eB})\tilde{\omega}_{c1}\tilde{\omega}_{c2}\right)}{(x-x_1)(x-x_2)}dx\right),\end{equation}
where \begin{equation} x_{1,2}=\left(\frac{\tilde\Delta}{\tilde{\omega}_{c1}+\tilde{\omega}_{c2}}-\frac{1}{2}-\frac{k^2_z}{2eB}\right)+ \frac{i\zeta_{m}(\tilde{\omega}_{c2}-\tilde{\omega}_{c1})}{2 \tilde{\omega}_{c1}\tilde{\omega}_{c2}}\pm \frac{i\sqrt{(\zeta_{m}(\tilde{\omega}_{c1}+\tilde{\omega}_{c2}))^2+4\tilde{\omega}_{c1}\tilde{\omega}_{c2}{\tilde\gamma}^2}}{2\tilde{\omega}_{c1}\tilde{\omega}_{c2}}.\end{equation}
Using the exponential form of  $\mr{sin}x$,
\begin{equation}\begin{split}\label{omega5}{\tilde{\Omega}}_{osc}&=\frac{DT}{\pi}\int_{-\infty}^{\infty}d{k_z}\Bigg( \sum_{l=1}^{\infty}\frac{1}{l}\sum_{{\zeta_m}}\left(-\frac{1}{\tilde{\omega}_{c1}\tilde{\omega}_{c2}}\right)\\ & \hspace{1.8cm}\times\frac{1}{2 i}\int_{0}^{\infty}\frac{{(e^{i 2\pi l x}-e^{-i 2\pi l x})}\left(i\zeta_m(\tilde{\omega}_{c2}-\tilde{\omega}_{c1})+\frac{2\tilde{\omega}_{c1}\tilde{\omega}_{c2}}{\tilde{\omega}_{c1}+\tilde{\omega}_{c2}}\tilde\Delta-2\tilde{\omega}_{c1}\tilde{\omega}_{c2} x-(1+\frac{k^2_z}{eB})\tilde{\omega}_{c1}\tilde{\omega}_{c2}\right)}{
(x-x_1)(x-x_2)}{dx}\Bigg).\end{split}
\end{equation}
We now want to compute the integral over $x$ using the residue theorem. Notice that the integrand contains the factors $e^{i2\pi lx}$ and $e^{-i2\pi lx}$, which have different convergence properties in the complex plane. Accordingly, we pick the appropriate contours: $C_1C_2C_3$ in the upper-half-plane for the former and $C_1'C_2'C_3'$ in the lower-half-plane for the latter, as shown in Fig.~\ref{path1}. Integrals over paths $C_2$ and $C_2'$ vanish according to Jordan's Lemma. On the other hand, integrals over the paths $C_3$ and $C_3'$ lead to nonoscillatory contributions which can be ignored. Thus, we eventually have
\begin{equation}
	\label{summation}
		{\tilde{\Omega}_{osc}} = \frac{DT}{\pi} \int_{-\infty}^{\infty}d{k_z}\sum_{l=1}^{\infty}\frac{1}{l}\sum_{{\zeta_m}}\left(-\frac{1}{\tilde{\omega}_{c1}\tilde{\omega}_{c2}}\right)\frac{1}{2i}(\text{summation over residues}).
\end{equation}
\begin{figure}[h]
	\centering
	\includegraphics[scale=0.25]{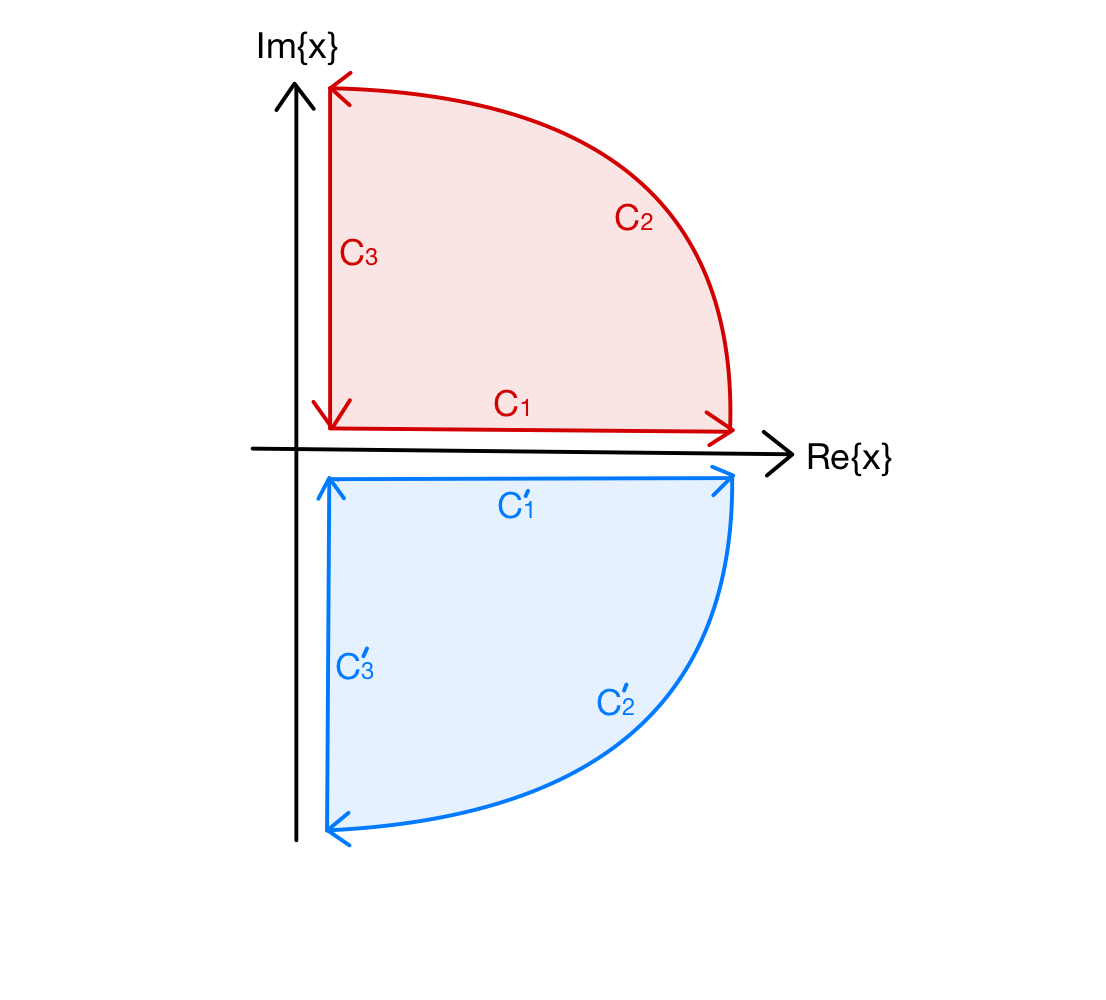}
	\caption{The contour used to evaluate the integral over $x$ in Eq.~(\ref{omega5}).}\label{path1}
\end{figure}
Because the location of the poles depends on the sign of $\zeta_m$, it is convenient to split the sum over $\zeta_m$ as $\sum_{\zeta_m}=\sum_{\zeta_m>0}+\sum_{\zeta_m<0}$ leading to $\tilde{\Omega}_{osc}={\tilde{\Omega}}^+_{osc}+{\tilde{\Omega}}^-_{osc}$. Equation~(\ref{summation}) then gives the following two contributions:
\begin{equation}\begin{split}
		{\tilde{\Omega}}^+_{osc}&=DT \int_{-\infty}^{\infty}d{k_z}\Bigg[\sum_{l=1}^{\infty}\frac{1}{l}\sum_{{\zeta_m}}\Bigg(e^{i2\pi l\left(\frac{\tilde\Delta}{\tilde{\omega}_{c1}+\tilde{\omega}_{c2}}-\frac{1}{2}-\frac{k^2_z}{2eB}\right)}e^{-2\pi l\left(\frac{\zeta_{m}(\tilde{\omega}_{c2}-\tilde{\omega}_{c1})}{2 \tilde{\omega}_{c1}\tilde{\omega}_{c2}}+ \frac{\sqrt{(\zeta_{m}(\tilde{\omega}_{c1}+\tilde{\omega}_{c2}))^2+4\tilde{\omega}_{c1}\tilde{\omega}_{c2}{\tilde\gamma}^2}}{2\tilde{\omega}_{c1}\tilde{\omega}_{c2}}\right)}\\ &\hspace{5cm} +e^{-i2\pi l\left(\frac{\tilde\Delta}{\tilde{\omega}_{c1}+\tilde{\omega}_{c2}}-\frac{1}{2}-\frac{k^2_z}{2eB}\right)}e^{2\pi l\left(\frac{\zeta_{m}(\tilde{\omega}_{c2}-\tilde{\omega}_{c1})}{2\tilde{\omega}_{c1}\tilde{\omega}_{c2}}- \frac{\sqrt{(\zeta_{m}(\tilde{\omega}_{c1}+\tilde{\omega}_{c2}))^2+4\tilde{\omega}_{c1}\tilde{\omega}_{c2}{\tilde\gamma}^2}}{2\tilde{\omega}_{c1}\tilde{\omega}_{c2}}\right)}\Bigg)\Bigg]
        \end{split}\end{equation}
and
\begin{equation}\begin{split}
		{\tilde{\Omega}}^-_{osc}&=DT \int_{-\infty}^{\infty}d{k_z}\Bigg[\sum_{l=1}^{\infty}\frac{1}{l}\sum_{{\zeta_m}}\Bigg(e^{i2\pi l\left(\frac{\tilde\Delta}{\tilde{\omega}_{c1}+\tilde{\omega}_{c2}}-\frac{1}{2}-\frac{k^2_z}{2eB}\right)}e^{-2\pi l\left(-\frac{\zeta_{m}(\tilde{\omega}_{c2}-\tilde{\omega}_{c1})}{2 \tilde{\omega}_{c1}\tilde{\omega}_{c2}}+ \frac{\sqrt{(\zeta_{m}(\tilde{\omega}_{c1}+\tilde{\omega}_{c2}))^2+4\tilde{\omega}_{c1}\tilde{\omega}_{c2}{\tilde\gamma}^2}}{2\tilde{\omega}_{c1}\tilde{\omega}_2}\right)}\\ &\hspace{5cm} +e^{-i2\pi l\left(\frac{\tilde\Delta}{\tilde{\omega}_{c1}+\tilde{\omega}_{c2}}-\frac{1}{2}-\frac{k^2_z}{2eB}\right)}e^{2\pi l\left(-\frac{\zeta_{m}(\tilde{\omega}_{c2}-\tilde{\omega}_{c1})}{2 \tilde{\omega}_{c1}\tilde{\omega}_{c2}}- \frac{\sqrt{(\zeta_{m}(\tilde{\omega}_{c1}+\tilde{\omega}_{c2}))^2+4\tilde{\omega}_{c1}\tilde{\omega}_{c2}{\tilde\gamma}^2}}{2\tilde{\omega}_{c1}\tilde{\omega}_{c2}}\right)}\Bigg)\Bigg].
\end{split}\end{equation} 
To carry out the integration over $k_z$, we use the method of steepest descents. We get the following expressions for $\tilde{\Omega}^{\pm}_{osc}$:
\begin{equation}
    \begin{split}
		\tilde\Omega^+_{osc}&=DT\sqrt{eB} \sum_{l=1}^{\infty}\frac{1}{l^{3/2}}\sum_{{\zeta_m}}\Bigg[e^{i2\pi l\left(\frac{\tilde\Delta}{\tilde{\omega}_{c1}+\tilde{\omega}_{c2}}-\frac{1}{2}-\frac{k^2_z}{2eB}\right)}e^{-\frac{i\pi}{4}}e^{-2\pi l\left(\frac{\zeta_{m}(\tilde{\omega}_{c2}-\tilde{\omega}_{c1})}{2 \tilde{\omega}_{c1}\tilde{\omega}_{c2}}+ \frac{\sqrt{(\zeta_{m}(\tilde{\omega}_{c1}+\tilde{\omega}_{c2}))^2+4\tilde{\omega}_{c1}\tilde{\omega}_{c2}{\tilde\gamma}^2}}{2\tilde{\omega}_{c1}\tilde{\omega}_{c2}}\right)}\\ &\hspace{5cm} +e^{-i2\pi l\left(\frac{\tilde\Delta}{\tilde{\omega}_{c1}+\tilde{\omega}_{c2}}-\frac{1}{2}-\frac{k^2_z}{2eB}\right)} e^{\frac{i\pi}{4}}
		e^{2\pi l\left(\frac{\zeta_{m}(\tilde{\omega}_{c2}-\tilde{\omega}_{c1})}{2 \tilde{\omega}_{c1}\tilde{\omega}_{c2}}- \frac{\sqrt{(\zeta_{m}(\tilde{\omega}_{c1}+\tilde{\omega}_{c2}))^2+4\tilde{\omega}_{c1}\tilde{\omega}_{c2}{\tilde\gamma}^2}}{2\tilde{\omega}_{c1}\tilde{\omega}_{c2}}\right)}\Bigg]
\end{split}
\end{equation}
and
\begin{equation}
    \begin{split}
		{\tilde{\Omega}}^-_{osc}&=DT\sqrt{eB} \sum_{l=1}^{\infty}\frac{1}{l^{3/2}}\sum_{{\zeta_m}}\Bigg[e^{i2\pi l\left(\frac{\tilde\Delta}{\tilde{\omega}_{c1}+\tilde{\omega}_{c2}}-\frac{1}{2}-\frac{k^2_z}{2eB}\right)}e^{-\frac{i\pi}{4}}e^{-2\pi l\left(-\frac{\zeta_{m}(\tilde{\omega}_{c2}-\tilde{\omega}_{c1})}{2 \tilde{\omega}_{c1}\tilde{\omega}_{c2}}+ \frac{\sqrt{(\zeta_{m}(\tilde{\omega}_{c1}+\tilde{\omega}_{c2}))^2+4\tilde{\omega}_{c1}\tilde{\omega}_{c2}{\tilde\gamma}^2}}{2\tilde{\omega}_{c1}\tilde{\omega}_{c2}}\right)}\\ &\hspace{5cm} +e^{-i2\pi l\left(\frac{\tilde\Delta}{\tilde{\omega}_{c1}+\tilde{\omega}_{c2}}-\frac{1}{2}-\frac{k^2_z}{2eB}\right)}e^{\frac{i\pi}{4}}e^{2\pi l\left(-\frac{\zeta_{m}(\tilde{\omega}_{c2}-\tilde{\omega}_{c1})}{2 \tilde{\omega}_{c1}\tilde{\omega}_{c2}}- \frac{\sqrt{(\zeta_{m}(\tilde{\omega}_{c1}+\tilde{\omega}_{c2}))^2+4\tilde{\omega}_{c1}\tilde{\omega}_{c2}{\tilde\gamma}^2}}{2\tilde{\omega}_{c1}\tilde{\omega}_{c2}}\right)}\Bigg].
	\end{split}
\end{equation}
Finally, adding ${\tilde{\Omega}}^{\pm}_{osc}$, we get the oscillatory grand potential,
\begin{equation}
    \tilde\Omega_{osc}=\frac{(eB)^{3/2}}{\pi^2}T\sum_{l=1}^{\infty}\frac{1}{l^{3/2}}\sum_{{\zeta_m>0}}  e^{-\frac{\pi l}{\tilde{\omega}_{c1}\tilde{\omega}_{c2}}\sqrt{\zeta_m^2(\tilde{\omega}_{c1}+\tilde{\omega}_{c2})^2+4\tilde{\omega}_{c1}\tilde{\omega}_{c2} \tilde\gamma^2}}
	\cosh\left[{\frac{\pi l\zeta_m (\tilde{\omega}_{c2}-\tilde{\omega}_{c1})}{\tilde{\omega}_{c1}\tilde{\omega}_{c2}}}\right]\mr{cos}\left[2\pi l\left(\frac{\tilde\Delta}{\tilde{\omega}_{c1}+\tilde{\omega}_{c2}}-\frac{1}{2}\right)-\frac{\pi}{4}\right].
	\label{finexp}
\end{equation}

\subsection{DEPENDENCE OF AMPLITUDE ON TEMPERATURE}

\begin{figure}
	\centering
	\includegraphics[width=\textwidth]{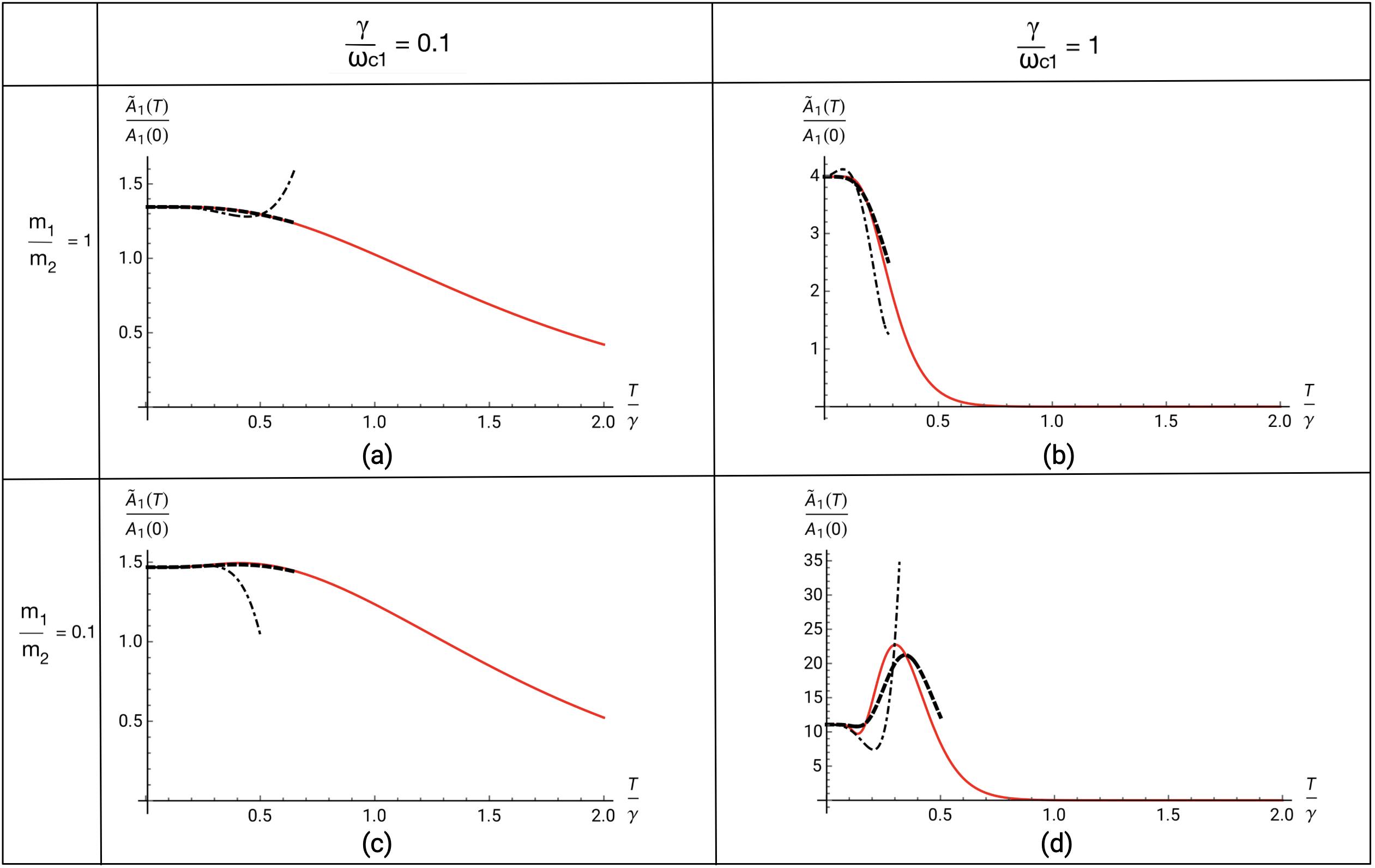}
	\caption{The plots are drawn for scenario R of Fig. 3 in the main text. The solid, red curve corresponds to Eq.~(\ref{ampgensupp}), the thick, dashed, black curve corresponds to Eq.~(\ref{tdepend}), and the thin dash-dotted, black curve corresponds to an expansion of Eq.~(\ref{tdepend}) in $T$ to $\mathcal{O}(T^5)$. In all the plots, $\gamma=1$.}\label{tdepen}\end{figure}
	
The amplitude of $\tilde\Omega_{\mr{osc}}$ can be read off from Eq.~(\ref{finexp}):
\begin{equation}
\tilde{A}_l(T)=
\frac{(eB)^{3/2}}{\pi^2l^{3/2}}T\sum\limits_{m=0}^{\infty}  e^{-\frac{\pi l  }{\tilde\omega_{c1}\tilde\omega_{c2}}\sqrt{(2m+1)^2\pi^2T^2(\tilde\omega_{c1}+\tilde\omega_{c2})^2+4\tilde\omega_{c1}\tilde\omega_{c2} \tilde\gamma^2}}
\cosh\left[{\frac{\pi l(2m+1)\pi T (\tilde\omega_{c2}-\tilde\omega_{c1})}{\tilde\omega_{c1}\tilde\omega_{c2}}}\right]\label{ampgensupp}.
\end{equation}
At low $T$, the summation over $m$ can be carried out using the Euler-Maclaurin formula:
\begin{equation}
\sum_{m=0}^{\infty}F(m)=\int_0^\infty F(r)dr +\frac{1}{2}
[F(\infty)+F(0)]+ \frac{1}{12}[F'(\infty)-F'(0)]+\cdots,\label{eulersupp}
\end{equation} 
where $F(m)$ is the summand in Eq.~(\ref{ampgensupp}). Making the change of variable $(2m+1)\pi T\rightarrow x$, which changes $F(m)\rightarrow f(x)=e^{-\frac{\pi l  }{\tilde\omega_{c1}\tilde\omega_{c2}}\sqrt{x^2(\tilde\omega_{c1}+\tilde\omega_{c2})^2+4\tilde\omega_{c1}\tilde\omega_{c2} \tilde\gamma^2}}
\cosh\left[{\frac{\pi k x (\tilde\omega_{c2}-\tilde\omega_{c1})}{\tilde\omega_{c1}\tilde\omega_{c2}}}\right]$, we have
\begin{equation}
    \tilde{A}_l(T)\approx \alpha\left[
    \frac{1}{2\pi} \int_{\pi T}^{\infty} f(x) dx + \frac{T}{2}f(\pi T)-\frac{T}{12}f'(\pi T)\right],\label{tdependone}
\end{equation}
where $\alpha=\frac{(eB)^{3/2}}{\pi^2l^{3/2}}$ and  we have used the fact that $f(x\rightarrow\infty)=f'(x\rightarrow\infty)=0$. Writing $\int_{\pi T}^\infty f(x)dx=\int_{0}^\infty f(x)dx-\int_0^{\pi T}f(x)dx$, and recognizing that the first integral is independent of $T$, we have 
\beq
\tilde{A}_l(T)\approx\tilde{A}_l(0)-\alpha\left[\frac{1}{2\pi} \int_{0}^{\pi T} f(x) dx -\frac{T}{2}f(\pi T)+\frac{T}{12}f'(\pi T)\right].\label{tdepend}
\eeq
Equation (\ref{tdepend}) reproduces the numerically obtained curve for $\tilde{A}_1(T)$ very well for $T\lesssim\gamma$ as seen in Fig.~\ref{tdepen} implying that the truncation used in the Euler-Maclaurin formula is justified. But, it is not in a closed analytical form, thanks to the integral. A possible way out is to further expand the integral in $T$. For consistency, we also need to expand the other two terms to the same order in $T$ even though they are algebraic expressions. Such a procedure does yield a closed analytical form for $\tilde{A}_l(T)$, but it is not very useful. We demonstrate this in Fig.~\ref{tdepen} where we consider an expansion of Eq.~(\ref{tdepend}) up to $\mathcal{O}(T^5)$. It is seen that the expression reproduces the numerically obtained curve only for $T<<\gamma$ but not for $T\lesssim\gamma$.

\subsection{Self-energy}
In this section, we will compute the self-energy for the Hamiltonian given in Eqs. (4) and (12) in the main text: 
\beq
H=\sum_{i,\mk}\ve_{i\mk}c_{i\mk}^\dagger c_{i\mk}+\sum_{\mk}\left(\gamma_\mk c_{1\mk}^\dagger c_{2\mk}+\mr{h.c.}\right)+\sum_{i,j,\mk,\mk^\prime,\mb{q}}V_{ij\mb{q}}	c^\dagger_{i\mk+\mb{q}}c^\dagger_{j\mk^{\prime}-\mb{q}}c_{j\mk^{\prime}}c_{i\mk}.
\label{orgham}
\eeq
The corresponding mean-field Hamiltonian can be written as follows:
\beq
H_{MF}=\sum_{i,\mk}\left[\ve_{i\mk}+\Sigma_{ii}(\mb{k})\right]c_{i\mk}^\dagger c_{i\mk}+\sum_{\mk}\left[\left(\gamma_\mk+\Sigma_{12}\right) c_{1\mk}^\dagger c_{2\mk}+\mr{h.c.}\right]
\eeq
where 
\begin{subequations}
\begin{align}
\Sigma_{ii}(\mb{k})&=\sum_{\mb{k}'}\left[\left(V_{ii\mb{0}}-V_{ii\mb{k}'-\mb{k}}\right)\langle c_{i\mb{k}'}^{\dagger}c_{i\mb{k}'}\rangle+V_{12\mb{0}}\langle c_{j\mb{k}'}^{\dagger}c_{j\mb{k}'}\rangle\right],\ \ (i\ne j),\\
\Sigma_{12}(\mb{k})&=-\sum_{\mb{k}'}V_{12\mb{k}'-\mb{k}}\langle c_{1\mb{k}'}^{\dagger}c_{2\mb{k}'}\rangle.
\end{align}
\end{subequations}
The averages over the ground state at $T=0$ can be calculated simply by going to the diagonal basis of the noninteracting part of the Hamiltonian:
\begin{subequations}
\begin{align}
H&=\sum_{i,\mk}\ve_{i\mk}c_{i\mk}^\dagger c_{i\mk}+\sum_{\mk}\left(\gamma_\mk c_{1\mk}^\dagger c_{2\mk}+\mr{h.c.}\right)\nonumber\\
&=\sum_{\mb{k}}\left[E_{-\mb{k}}d_{-\mb{k}}^\dagger d_{-\mb{k}}+E_{+\mb{k}}d_{+\mb{k}}^\dagger d_{+\mb{k}}\right],
\end{align}
\end{subequations} 
where $E_{\pm\mb{k}'}=\frac{1}{2}\Big[(\ve_{1\mb{k}'}+\ve_{2\mb{k}'})\pm\sqrt{(\ve_{1\mb{k}'}-\ve_{2\mb{k}'})^2+4\gamma^2}\Big]$, with $-,+$ referring to the hybridized valence and conduction bands, respectively. The basis transformation is given by the matrix:
\beq
\begin{bmatrix}
c_{1\mb{k}}\\c_{2\mb{k}}
\end{bmatrix}
=
\begin{bmatrix}
u_{11}&u_{12} \\ u_{21}&u_{22}
	\end{bmatrix}
	\begin{bmatrix}
d_{-\mb{k}}\\d_{+\mb{k}}
\end{bmatrix},
\eeq
where $u_{ij}$ are derived from the eigenvectors as usual. To calculate $\langle c_{i\mb{k}'}^{\dagger}c_{i\mb{k}'}\rangle$, we can simply ignore $\gamma$ and write $\langle c_{i\mb{k}'}^{\dagger}c_{i\mb{k}'}\rangle=n_F(\xi_{i\mb{k}'})$, where $\xi_{i\mb{k}'}=\ve_{i\mb{k}'}-\mu$ and $n_F(x)$ is the Fermi function. To calculate $\langle c_{1\mb{k}'}^{\dagger}c_{2\mb{k}'}\rangle$, we go to the new basis and write $\langle c_{1\mb{k}'}^{\dagger}c_{2\mb{k}'}\rangle=\langle(u_{11}^\ast d_{-\mb{k}'}^\dagger+u_{12}^\ast d_{+\mb{k}'}^\dagger)(u_{21} d_{-\mb{k}'}+u_{22} d_{+\mb{k}'})\rangle$. Noting that $\langle d_{\pm\mb{k}'}^\dagger d_{\pm\mb{k}'}\rangle=n_F(E_{\pm\mb{k}'})$ and $\langle d_{\pm\mb{k}'}^\dagger d_{\mp\mb{k}'}\rangle=0$, together with $n_F(E_{+\mb{k}'})=0$ at $T=0$, we finally arrive at the following expressions:

\begin{subequations}
\begin{align}
\Sigma_{ii}(\mb{k})&\approx\sum_{\mb{k}'}\left[\left(V_{ii\mb{0}}-V_{ii\mb{k}'-\mb{k}}\right)n_F(\xi_{i\mb{k}'})+V_{12\mb{0}}n_F(\xi_{j\mb{k}'})\right],\\
\Sigma_{12}(\mb{k})&=\gamma\sum_{\mb{k}'}V_{12\mb{k}'-\mb{k}}\frac{n_F(E_{-\mb{k}'})}{(E_{+\mb{k}'}-E_{-\mb{k}'})},
\end{align}
\end{subequations} 
\end{widetext}
\bibliography{qoscreflibrary}

\end{document}